\documentclass[preprint,showpacs,a4,prl]{revtex4} 
\usepackage{graphicx}%
\usepackage{dcolumn} 
\usepackage{bm} 
\makeatletter \def\btt#1{\texttt{\@backslashchar#1}}%
\DeclareRobustCommand\bblash{\btt{\@backslashchar}}%
\makeatother  
\begin{document}  
\preprint{cation-deficient.TEX}  
\title{Spin-polarized oxygen hole states in cation deficient La$_{1-x}$Ca$_x$MnO$_{3+\delta}$ }
\author{G. Papavassiliou$^1$, M. Pissas$^1$, M. Belesi$^1$, M. Fardis$^1$, J. P. Ansermet$^{2,3}$, D. Carlier$^{3}$, C. Dimitropoulos$^{3}$, and J. Dolinsek$^4$} 
\affiliation{$^1$Institute of Materials Science, NCSR,  Demokritos, 153 10 Aghia Paraskevi, Athens, Greece\\
$^2$Dept. of Physics, University of Illinois, Urbana, Illinois 1801, USA\\
$^3$Institut de Physique Exprimentale, EPFL-PH-Ecublens, 1015-Lausanne, Switzerland\\
$^4$Josef Stefan Institute, Jamova 39, 61111 Ljubljana, Slovenia
} 
\date{\today }   
\begin{abstract} 
When holes are doped into a Mott-Hubbard type insulator, like lightly doped manganites of the La$_{1-x}$Ca$_x$MnO$_3$ family, the cooperative Jahn-Teller distortions and the appearance of orbital ordering require an arrangement of Mn$^{3+}$/Mn$^{4+}$ for the establishment of the insulating canted antiferromagnetic (for $x\leq 0.1$), or of the insulating ferromagnetic (for $0.1\leq x\leq 0.2$) ground state. In the present work we provide NMR evidence about a novel and at the same time puzzling effect in La$_{1-x}$Ca$_x$MnO$_{3+\delta}$ systems with cation deficience. We show that in the low Ca-doping regime, these systems exhibit a very strong hyperfine field at certain La nuclear sites, which is not present in the stoichiometric compounds. Comparison of our NMR results with recent x-ray absorption data at the Mn $K$ edge, suggests the formation of a spin-polarized hole arrangement on the $2p$ oxygen orbitals as the origin of this effect.  
\end{abstract} 
\pacs{75.70.Pa., 76.20.+q, 75.30.Et, 75.60.Ch} 
\maketitle

The presence of oxygen excess in low doped colossal magnetoresistive manganites, like La$_{1-x}$Ca$_x$MnO$_{3+\delta}$ (LCMO$_\delta$) for $x\leq 0.20$ alters dramatically the ground state of these systems: The cooperative Jahn-Teller (JT) distortions smooth out, as implied by the disappearance of the characteristic JT split in the x-ray diffraction peaks, and the low temperature canted antiferromagnetic state of the stoichiometric compounds (for $x\leq 0.1$) becomes ferromagnetic insulating, while $T_c$ increases substantially \cite{Subias98,Huang98}. This effect is usually explained as consequence of inducing a higher fractional hole concentration $c_h$ than the nominal one, which makes the system equivalent to a system with higher doping $x$. Specifically, in addition to holes from the divalent cation replacement, excess of oxygen is considered to create equal number of La and Mn vacancies, while converting Mn ions to the $(4+)$ state, so that $c_h=x+2\delta$ \cite{Subias98,Hundley97}. The present work unravels a novel and at the same time puzzling feature, observed in low Ca doped cation deficient LCMO$_\delta$.  By applying  $^{55}$Mn and $^{139}$La NMR spectroscopy on a series of stoichiometric ($\delta=0$) and cation deficient, non-stoichiometric ($\delta \neq 0$), LCMO$_\delta$ systems, we demonstrate that oxygen excess gives rise to a peculiar magnetoelectronic arrangement for $0\leq x\leq 0.20$. At temperatures lower than $40$K the FM insulating phase breaks spontaneously in two strongly intermixed subphases: (i) A FM insulating phase, where the expected Mn$^{3+,4+}$ electron states are experimentally resolved with $^{55}$Mn NMR, and (ii) a FM overstructure, which is characterized by a surprisingly high transfered hyperfine field at the position of the La nuclei, and a broad distribution of frequencies in the $^{55}$Mn NMR. By increasing Ca doping this second phase component decreases rapidly, and becomes unimportant for $x\geq 0.20$. It also reduces drastically by increasing temperature, or by applying an external magnetic field. Comparison with Mn $K$ edge x-ray absorption spectroscopy (XAS) data implies that for $x\leq 0.20$, the low temperature state of non-stoichiometric compounds contains an arrangement of spin-polarized holes at the oxygen sites, aparently in hybridized Mn$(3d)$-O$(2p)$ states with a strong O$(2p)$ character, which are imbibed into a FM insulating matrix resembling the FM insulating phase of the stoichiometric compounds. 
 
The idea about the formation of spin-polarized O$(2p)$ holes in the manganites is not new. On the basis of oxygen K-edge electron energy loss spectroscopy, Ju et al. \cite{Ju97} concluded that charge carriers in La$_{1-x}$Sr$_x$MnO$_3$ (LSMO) thin films reside on the O$(2p)$ orbitals, while by using photoemission and XAS techniques in Sr and Ca doped manganites, a number of investigators \cite{Abbate92,Chainani93,Saitoh95,Croft97,Qian03} have shown strong Mn$(3d)$-O$(2p)$ hybridization with strong O$(2p)$ character of the doped holes. At the same time, polarized neutron scattering studies in LSMO \cite{Pierre98} and the layered La$_{1.2}$Sr$_{1.8}$Mn$_2$O$_7$ \cite{Argyriou02} have confirmed the presence of positive spin density on the oxygens. From the theoretical point of view, ab initio Hartree-Fock calculations by Freyria-Fava \cite{Fava97} and Pickett and Singh \cite{Pickett96} predicted positive spin-density on the oxygen atoms in perovskite manganites. Other Hartree-Fock calculation for La$_{0.5}$Ca$_{0.5}$MnO$_3$, indicate the formation of a low temperature ordering of spin-polarized oxygen holes in  a charge density wave \cite{Ferrari03}. Contrary to the above, there is a great number of experimental and theoretical work in the literature, which indicates that magnetic ordering in low doped LCMO and LSMO is dictated by the cooperative Jahn-Teller distortions and the respective orbital ordering, with Mn$^{4+}$/Mn$^{3+}$ charge mixtures and unpolarized O$^{2-}$ ions \cite{Tokura00,Dagotto02}. At the same time, low temperature neutron scattering data in lightly doped LCMO \cite{Hennion01} and LSMO \cite{Inami99,Yamada00} show a complex picture of the spin and charge configuration. The situation becomes even more complicated as there are numerous reports about the formation of a glassy phase component and partial disorder in these systems at low temperature 
\cite{Dai00,Laiho01}. 
    
In this respect, NMR which probes the local environment of the resonating nuclei is a very promissing probe in identifying the ground state at a local level, even at length-scales much below the limit of x-ray and neutron scattering techniques.  In case of magnetic materials as manganites, NMR in zero external magnetic field probes the local magnetic environment of the resonating nuclei through the hyperfine field, $B_{hf}=\frac 1{\gamma \hbar }A\left\langle S\right\rangle$, where $A$ and $\left\langle S\right\rangle $ are the hyperfine coupling constant and the average electronic spin, respectively. According to this formula, $^{55}$Mn NMR in LCMO probes the electron spin state of single Mn ions and therefore is possible to detect the different Mn charge states, i.e. localized Mn$^{3+,4+}$, as well as intermediate valence states. Previous works in manganese perovskites have shown that at low temperatures NMR signals from Mn$^{4+}$ and Mn$^{3+}$ charge states have peaks at frequencies $\approx 320-330$ MHz and $\approx 420-430$ MHz, respectively, while signals from FM metallic regions are located at intermediate frequencies, due to the fast hole transfer between the Mn$^{3+,4+}$ states \cite{Kapusta99,Belesi01,Alodi02,Savosta03}.  On the other hand, $^{139}$La NMR (the La ion has total electron spin $S=0$), probes the average spin state of the surrounding Mn octant through transferred hyperfine interactions, and therefore is very sensitive to changes of the local Mn spin configuration. Symmetry arguments indicate that the hyperfine field $B_{hf}(La)$ varies from zero value for collinearly AFM ordered Mn octants, to a maximum value for collinearly FM ordered Mn octants. Recent studies have shown that $B_{hf}(La)$ is not sensitive in changes of the $e_g$ states population \cite{Papavassiliou99}. This has been explained as showing that $B_{hf}(La)$ arises indirectly via $\pi $ type overlapping between the Mn $\left| 3d_{xy}\right\rangle $, $ \left| 3d_{yz}\right\rangle $, $\left| 3d_{zx}\right\rangle $, and the oxygen $\left| 2p_\pi \right\rangle $ wave functions, in conjunction with $\sigma $ bonding of the oxygen with the $\left| sp^3\right\rangle $ hybrid states of the La$^{3+}$ ion \cite{Papavassiliou99}. We stress that since $B_{hf}(La)$ is mediated via the oxygens, spin-polarized oxygen sites - if present - are expected to alter substantially $B_{hf}(La)$ at the closest neighbors. 

LCMO$_\delta$ samples were prepared by thoroughly mixing high-purity stoichiometric amounts of La$_2$O$_3$, MnO$_2$, and CaCO$_3$. The non-stoichiometric compounds were obtained as prepared, following the synthetic route previously reported in ref. \cite{Papavassiliou99}. X-ray powder diffraction showed single phase patterns for all compounds. The refinement indicated the presence of La and Mn vacancies, corresponding to $\delta \approx 0.08-0.1$. The stoichiometric ($\delta =0$) compounds were obtained by annealing as prepared compounds in ultra pure He atmosphere at $1000$K. Zero external field $^{139}$La and $^{55}$Mn NMR signals were acquired on a home built spectrometer equipped with an Oxford cryostat, by applying a two pulse spin-echo technique. The NMR line shapes were then obtained by measuring the integrated intensity of the spin-echo as a function of frequency. 

\begin{figure}[tbp] 
\centering 
\includegraphics[angle=0,width=7cm]{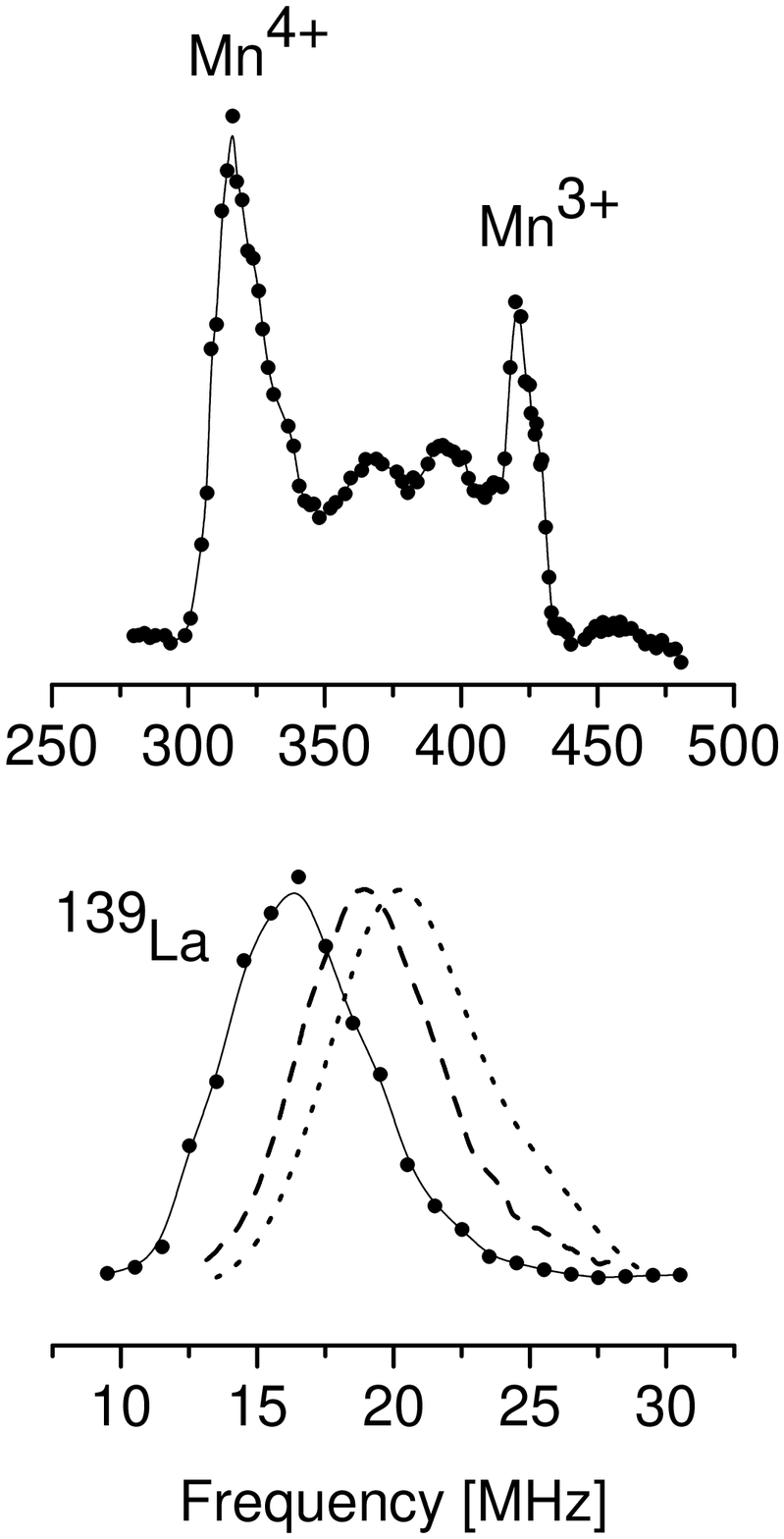} 
\caption{$^{139}$La (bottom) and $^{55}$Mn (top) NMR of stoichiometric LCMO$_{\delta=0}$ $x=0.125$ at $5$K and $15$K respectively. The dashed line is  $^{139}$La NMR signal for $x=0.25$, and the dotted line for $x=0.33$.}
\label{Fig1} 
\end{figure}  

Figure \ref{Fig1} shows $^{55}$Mn and $^{139}$La NMR line shapes of a stoichiometric LCMO$_{\delta=0}$ system with $x=0.125$. The $^{55}$Mn spectrum is characterized by two well defined "edge" peaks, which correspond to the Mn$^{3+,4+}$ valence states, respectively, in accordance with the literature. In addition, extra peaks are observed at intermediate frequencies, which might indicate the presence of Mn ions with valence state between $3+$ and $4+$. Similar results were recently reported with $^{55}$Mn NMR measurements on LCMO $x=0.1$ \cite{Algarabel03}, and LSMO $x=0.16$ \cite{Savosta03a}. The simplest neutron diffraction picture corresponding to an arrangement containing intermediate valence states is that of Inami \cite{Inami99}, where hole-rich layers, stacked along [001], with periodicity $4a_0$, contain $1/2$Mn$^{4+}$+$1/2$Mn$^{3+}$ charge states, while Yamada et al. \cite{Yamada00} have proposed an even more complex arrangement. On the contrary, $^{139}$La NMR spectrum consists of a featureless single Gaussian line shape, providing a further proof that $^{139}$ La NMR signal is insensitive to the population of the $e_g$ hole states. It is also observed that by increasing doping the signal peak shifts gradually to higher frequencies, as shown in the lower pannel of Figure \ref{Fig1}. Considering that the unit cell volume decreases monotonously with Ca doping in stoichiometric compounds, it may be concluded that the frequency increase should be assigned to an increase of the wave function overlaping, i.e. an increase of the hyperfine coupling constant $A$ \cite{Papavassiliou99}. 

\begin{figure}[tbp] 
\centering 
\includegraphics[angle=0,width=7cm]{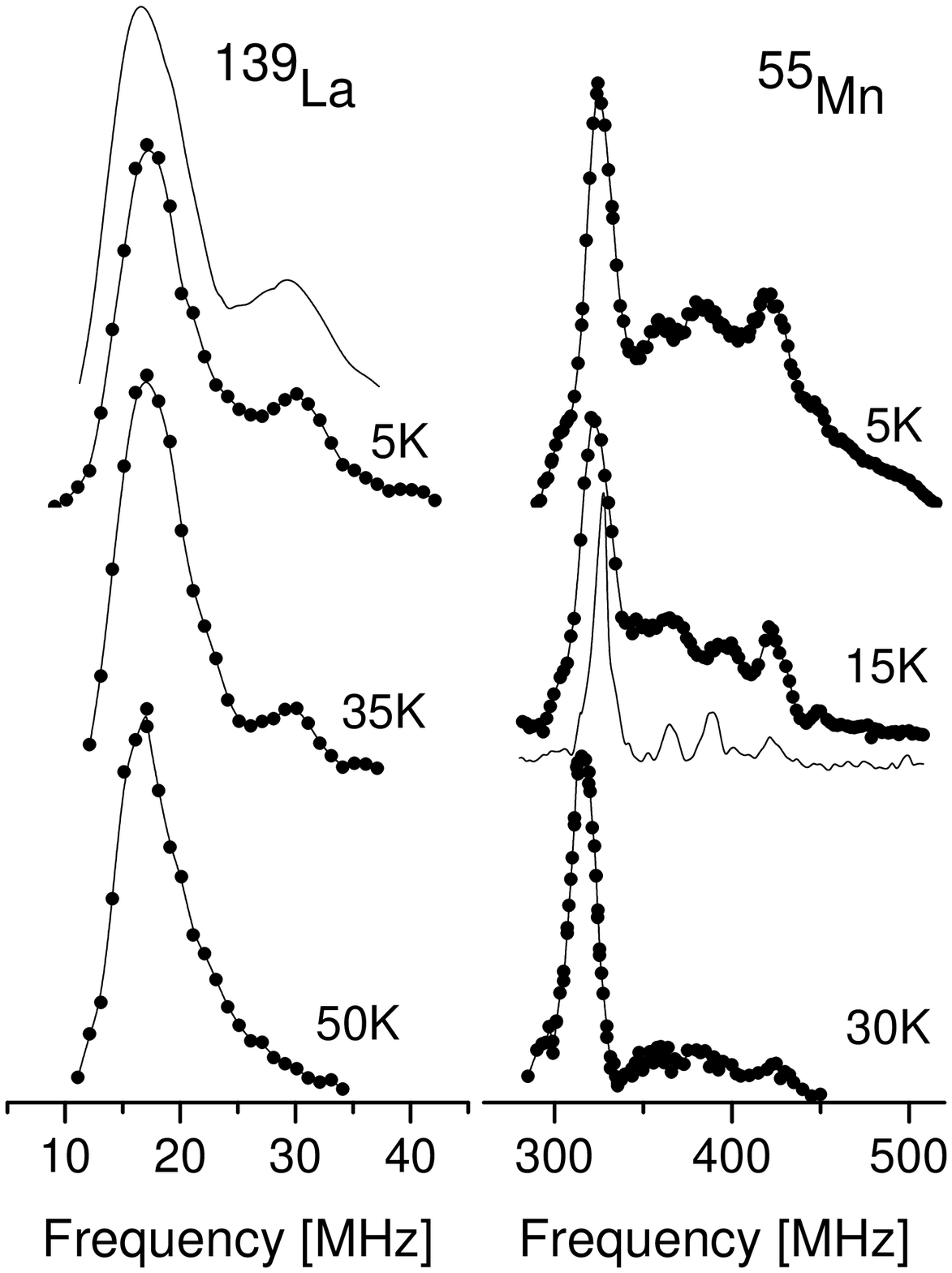} 
\caption{$^{139}$La and $^{55}$Mn NMR of non-stoichiometric LCMO$_{\delta \neq 0}$ $x=0.125$, in zero external magnetic field as a function of temperature. The solid line at the top of the $^{139}$La spectra is the spectrum for non-stoichiometric $x=0.075$ at $5$K. The solid line of the $^{55}$Mn spectra at $15$K is taken in magnetic field of $0.6$ Tesla.}
\label{Fig2} 
\end{figure}

\begin{figure}[tbp] 
\centering 
\includegraphics[angle=0,width=7cm]{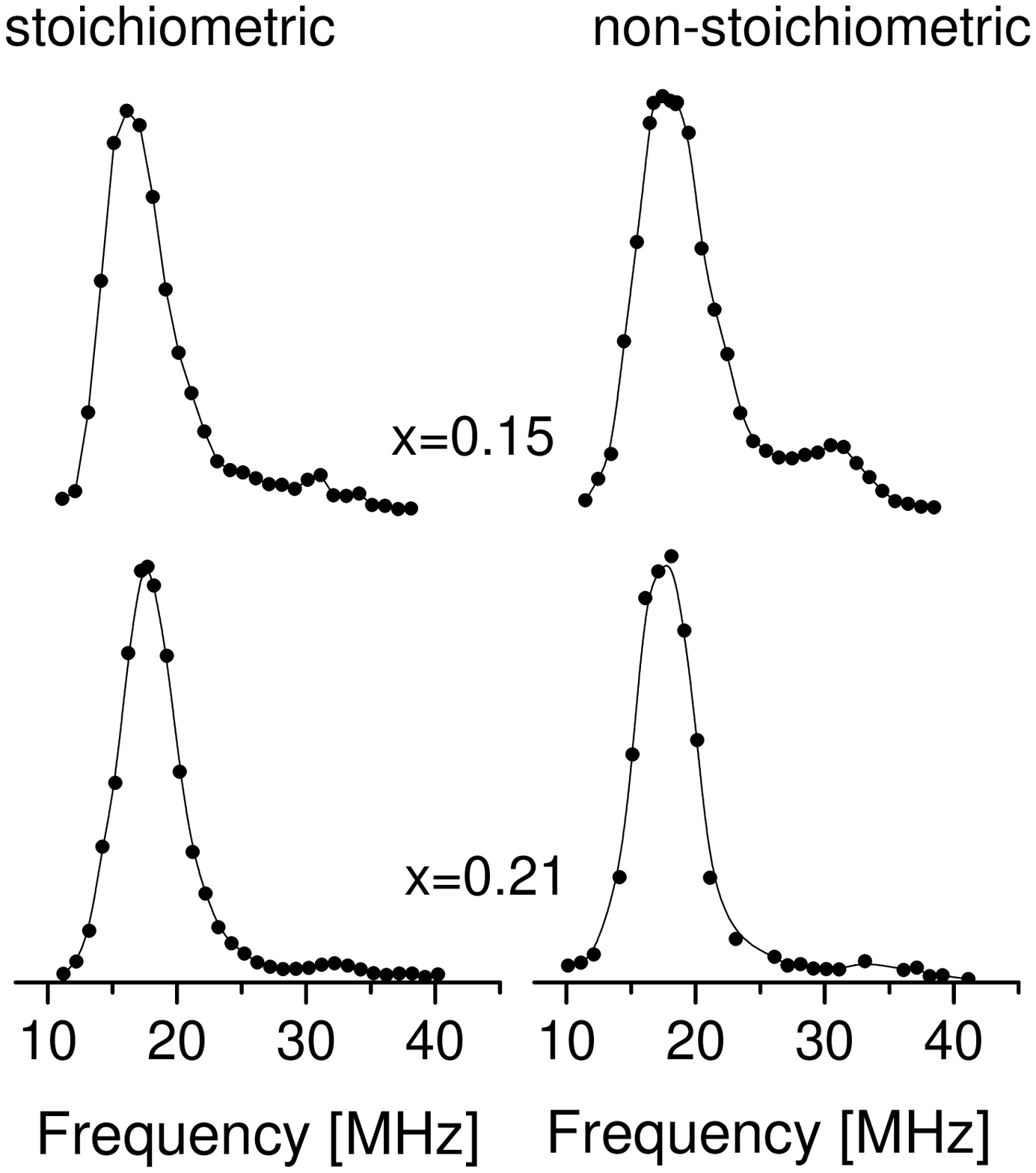} 
\caption{$^{139}$La NMR of stoichiometric LCMO$_{\delta =0}$ and non-stoichiometric LCMO$_{\delta \neq 0}$ $x=0.15$ and $0.21$, in zero external magnetic field at $T=5$K.}
\label{Fig3} 
\end{figure}

It is thus surprising that at low temperatures, non-stoichiometric LCMO$_{\delta \neq 0}$ compounds with $x\leq 0.20$, exhibit an extra $^{139}$La NMR signal with frequency $\approx 30$ MHz (Figure \ref{Fig2}), which exceeds by far the frequency $\approx 20.3$ MHz of fully polarized Mn octants in the FM metallic $x=0.33$. This secondary high frequency peak reduces drastically by increasing doping, and becomes negligibly small for $x\geq 0.20$, as shown in Figure \ref{Fig3}. We also notice that the main signal component at $\approx 17.5$ MHz remains almost unshifted for $x\leq 0.20$ in the non-stoichiometric compounds. According to neutron scattering and X-ray diffraction experiments  \cite{Huang98,Croft97} the unit cell volume in non-stoichiometric compounds varies only slightly ($V\approx 234$ $\AA ^3$ for $x\leq 0.15$ at $T=10$K). This keeps the hyperfine coupling constant $A$ almost invariant in the low doping regime. Besides, the increase in the unit cell volume of the stoichiometric $x=0.125$ compound ($V\approx 237$ $\AA ^3$ \cite{Huang98}) is accompanied by only a slight decrease in the frequency of the main peak. On the basis of these observations the high frequency $^{139}$La signal can not be ascribed to variation of the Mn charge states, or any kind of changes in the unit cell geometry, but as discussed below to extra spin polarization on the oxygen sites. A similar behaviour is shown by the $^{55}$Mn NMR spectra in Figure \ref{Fig2}. Signals from non-stoichiometric compounds appear to consist of two overlapping signal components: a signal similar to the one from the stoichiometric compound, and a shapeless broad line with a maximum at frequency $\approx 380-390$ MHz. The frequency of this second signal is between the frequencies of the Mn$^{4+,3+}$ peaks,  and may be assigned to the extra holes  residing on the Mn-O orbitals, which give rise to a Mn spin state with effective spin $S<2$. By applying an external magnetic field or by increasing temperature the additional signal decreases rapidly and practically only the Mn$^{4+}$ signal is observable (Figure \ref{Fig2}).

\begin{figure}[tbp] 
\centering 
\includegraphics[angle=0,width=7cm]{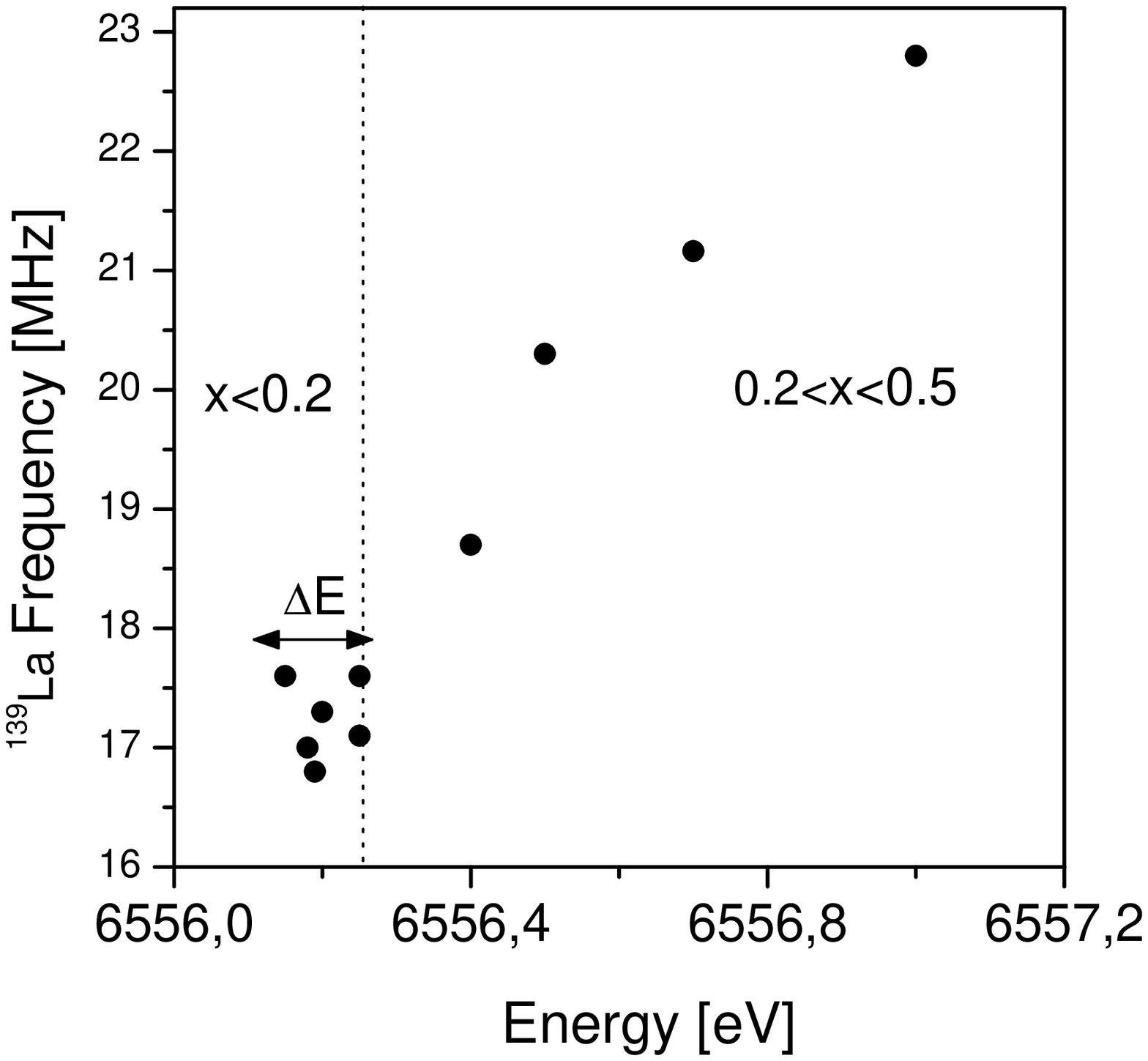} 
\caption{$^{139}$La NMR frequency as a function of the Mn K-edge peak position from XAS measurements \cite{Croft97} in non-stoichiometric LCMO$_{\delta \neq 0}$ compounds.}
\label{Fig4} 

\end{figure}

We anticipate that the extra NMR signal in lightly doped and cation deficient systems is a consequence of the fact that the ground state in these compounds is dictated by charge transfer and strong hybridization of the Mn-O bond \cite{Abbate92,Fava97,Pickett96}. Early in the 50's, Goodenough \cite{Goodenough55} has addressed this problem by inventing the concept of "semicovalent exchange" to describe the Mn$(3d)$-O$(2p)$ hybridization. According to this phenomenological approach, covalent bonding of oxygens with only one Mn ion, in conjunction with Hund's coupling, lead to negative spin polarization and hole-type charge transfer on the oxygen ions. However, polarized neutron experiments and elaborate ab initio calculations have recently shown that spin polarization on oxygens is positive instead of negative \cite{Pierre98,Fava97,Pickett96},  and aspherical \cite{Pierre98}, reminding that of $p_\pi $ orbitals. {\it It is such oxygen spin polarization, which can be only responsible for the $30$ MHz $^{139}$La NMR signal}. This opinion is in excellent agreement with recent Mn-K edge XAS results in cation deficient LCMO$_{\delta \neq 0}$ samples \cite{Croft97}. According to these measurements, $e_g$ hole formation opens at Ca doping $\approx 0.20$, while at lower doping holes are considered to possess a significant O$(2p)$ character. The nice correlation of the $^{139}$La NMR results with the Mn K-edge XAS data is clearly seen in Figure \ref{Fig4}. For doping $x\leq 0.20$, i.e. in the doping range where the high frequency extra peak at $\approx 30$ MHz is detectable, the main $^{139}$La NMR peak remains almost unshifted at frequency $\approx 17.5$ MHz. At the same time the energy of the Mn K-edge peak is randomly distributed in the range $\Delta E\approx 6556,1-6556,3$ $eV$. However, for $x\geq 0.20$ there is  a substantial linear decrease of the unit cell volume with $x$, accompanied by an increase of the NMR frequency, and a remarkable increase of the XAS peak energy. As pointed in ref. \cite{Croft97} this energy increase is a sign of the formation of Mn$^{4+}$ ionic sites.  It may thus be inferred that our NMR results indicate a predominant O$(2p)$ character for holes in cation deficient compounds with $x\leq 0.20$. Further support to this arguments is given by Electron-Energy-Loss measurements of the $O_K$ edge in non-stoichiometric LaMnO$_{3+\delta }$ \cite{Ju97}. These measurements show a strong oxygen prepeak, which illustrates a significant $O(2p)$ hole density in the non-stoichiometric compounds, that is not present in the annealed compounds. What is worth to notice is that the ground state of the cation deficient compounds is highly inhomogeneous, as in these systems the charge and spin ordered state of the stoichiometric compounds appears to coexists with a spin-polarized hole density on oxygen sites. In terms of the Mott-Hubbard energy $U$ (i.e the energy needed to transfer one $e_g$ electron from one Mn ion to the next) and the charge transfer energy $\Delta $ (the energy needed to transfer one electron from the ligand band to the Mn site), we may conclude that $\Delta \approx U$, and the holes will go to states of heavily mixed character, thus giving rise at low temperatures to the spin-polarized oxygen hole overstructure. We notice that the recent observation of FM metallic walls in the FM insulating matrix for $x\geq 0.175$ might indicate a gradual "metallization" of the oxygen-hole states by increasing doping \cite{Papavassiliou03}.        

In conclusion, $^{139}$La and $^{55}$Mn NMR line shape measurements on stoichiometric ($\delta =0$) and non-stoichiometric ($\delta \neq 0$) LCMO$_\delta $ compounds, clearly show that extra holes from excess of oxygen in non-stoichiometric compounds does not contribute to the formal Mn$^{3+}$ to Mn$^{4+}$ configurational change in the low doping regime: strong hybridization of the Mn-O orbitals gives rise to phase separation in regions with $d^3L$ charge states, $L$ denoting a hole in the O$(2p)$ state, imbibed into a matrix with d$^{4+}\rightarrow $d$^{3+}$ charge fluctuations. At temperatures lower than $\approx 40$K, this hybridized high spin state orders into a charge density wave with FM order. 

This work has been partially supported by the Greek-Slovenian cooperation project No. 2495.

\end{document}